# Outdoor Crowd Flow Estimation Using RSRP from Commercial LTE Base Station: A Field Study

Kaisei Higeta
Faculty of Science and Technology
Sophia University
Tokyo, Japan
k-higeta-0v3@eagle.sophia.ac.jp

Masakatsu Ogawa
Faculty of Science and Technology
Sophia University
Tokyo, Japan
m-ogawa@sophia.ac.jp

Tomoki Murakami
Access Network Service
Laboratories
NTT, Inc.
Kanagawa, Japan
tomoki.murakami@ntt.com

Kazuya Ohara
Access Network Service
Laboratories
NTT, Inc.
Kanagawa, Japan
kazuya.ohara@ntt.com

Shinya Otsuki
Access Network Service
Laboratories
NTT, Inc.
Kanagawa, Japan
shinya.otsuki@ntt.com

*Abstract*— With the advent of the 6G era, Integrated Sensing and Communications (ISAC) has attracted increasing attention. One representative of use cases is crowd flow estimation on outdoor streets. However, most existing studies have focused on indoor environments or vehicles, and demonstrations of outdoor crowd flow estimation using commercial LTE base station remain limited. This study addresses this use case and proposes an analysis of a crowd flow estimation method using Reference Signal Received Power (RSRP) obtained from a commercial LTE base station. Specifically, pedestrian counts derived from a camera-based object recognition algorithm were associated with the variance of RSRP. The features obtained from the variance were quantitatively evaluated by combining a CatBoost regression model with SHapley Additive exPlanations (SHAP) analysis. Through this investigation, we clarified that an optimal variance window size for RSRP is 0.1–0.2 seconds and that enlarging the counting area increased the features obtained from the variance of RSRP, for machine learning. Consequently, this study is the first to quantitatively demonstrate the effectiveness of outdoor crowd flow estimation using commercial LTE, while also revealing the characteristic behavior of variance window size and counting area size in feature design.

## I. Introduction

In recent years, environmental sensing utilizing wireless communication signals has been actively studied. In particular, device-free sensing with Wi-Fi has been applied to human activity recognition [1], gesture estimation [2], crowd counting, and occupancy estimation [3], producing numerous research outcomes. However, since Wi-Fi is primarily designed for indoor use, its coverage and stability are limited, restricting its applicability to large-scale outdoor environments.

3GPP has discussed Integrated Sensing and Communications (ISAC), which leverages frequency bands used for communication for both communication and sensing, and it has attracted growing attention toward the 6G era [4][5]. In particular, ISAC is expected to serve as a fundamental technology that enables the deployment of new services by realizing simultaneous communication and sensing using existing commercial communication infrastructure [6].

Various studies have also reported the use of wireless sensing using LTE signals, primarily in indoor environments. For instance, gesture recognition using femtocell or commercial 4G microcell base stations [7][8], crowd density estimation with LTE synchronization signals transmitted by eNodeB [9], and indoor localization with simulated base stations [10] have been explored. For outdoor environments, position estimation in scenarios without explicit base station type [11] and vehicle speed or traffic flow estimation [12] have been reported. Furthermore, as an ISAC use case, another study demonstrated the effectiveness of CSI-based autonomous mobile robot (AMR) crossing detection in a smart factory scenario using an experimental base station [13]. Nevertheless, these prior studies have been mainly confined to controlled laboratory or indoor environments, or to outdoor scenarios limited to position estimation and vehicle sensing. Experimental demonstrations of crowd flow estimation in outdoor environments using operational commercial LTE base stations remain extremely scarce. Moreover, from the perspective of feature design in machine learning, the effects of factors such as the variance window size of RSRP and the size of counting areas on estimation performance have not been sufficiently investigated.

To explore the feasible use case, this study conducted a proof-of-concept experiment on crowd flow estimation in an outdoor street environment by measuring Reference Signal Received Power (RSRP) obtained from commercial LTE base stations using the open-source platform OpenAirInterface (OAI) [14], which enables cost-effective LTE/5G deployment. In particular, we focused on the effects of variance window size and counting area size, and systematically clarified the factors contributing to estimation accuracy by evaluating feature contributions using a CatBoost regression model and SHAP analysis. The results of this study provide, for the first time, a quantitative demonstration of the effectiveness of outdoor crowd flow estimation using commercial LTE, offering fundamental insights for future 5G/6G ISAC applications.

In summary, the contributions of this study are twofold. First, it presents the first quantitative field demonstration of outdoor crowd flow estimation using RSRP obtained from an operational commercial LTE base station. Second, it systematically reveals how the RSRP variance window size and the pedestrian counting area size influence estimation performance, thereby clarifying the characteristic behavior of these features in ISAC-based urban sensing using existing LTE infrastructure.

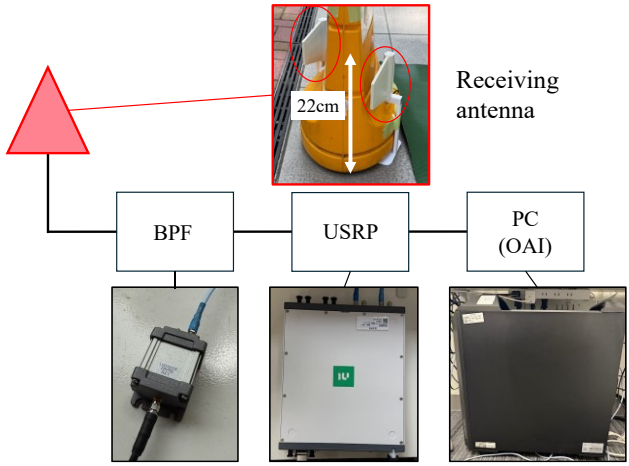

Fig.1  Measurement system.

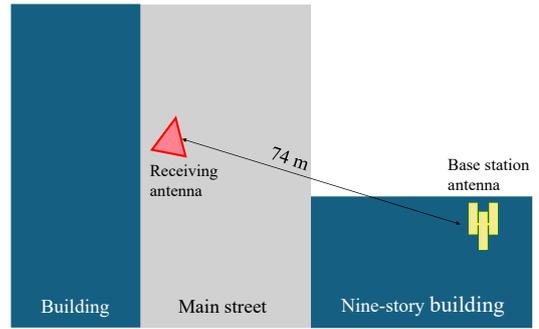

Fig.3  Geometric configuration between the base station antenna and the reception antenna.

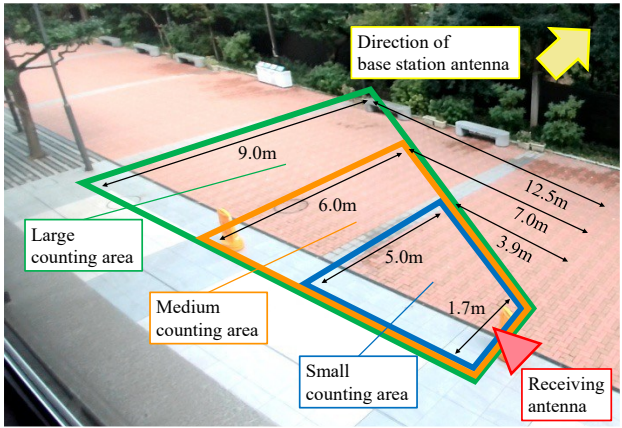

Fig.2  Measurement environment.

## II. EXPERIMENTAL OVERVIEW

In this study, we conducted a crowd estimation experiment on the main street of a university campus. Fig. 1 shows the configuration of the measurement system. This system consists of a receiving antenna, a band-pass filter (BPF), a Universal Software Radio Peripheral (USRP), and a PC of OpenAirInterface (OAI). The receiving antenna of this system receives the Cell-specific Reference Signal transmitted from the commercial base station, and the RSRP was calculated from the received power of this signal. RSRP was measured at an interval of 0.01 seconds.

The measurement environment is shown in Fig. 2. A directional antenna (Siretta OSCAR20A, length: 22 cm) was installed as the receiving antenna for RSRP measurements. To measure the number of pedestrians using a camera-based object recognition algorithm, three counting areas of different sizes were defined along the main street, with areas of 13.1 m² (Small area), 27.0 m² (Medium area), and 66.9 m² (Large area), respectively. The frequency band used for measurements was 2.1 GHz. Since the operator does not open the exact location of the commercial base station, we investigated the maximum RSRP measured by iPhone's field test mode using the Cell ID of the target base station. As a result, we confirmed that the base station's antenna corresponding to the target Cell ID was installed on the rooftop of a nine-story building. According to the Ministry of Land, Infrastructure, Transport and Tourism's PLATEAU dataset [15], the building height is 50 m, and the horizontal distance from the receiving antenna was estimated to be 55 m using Google Maps. Consequently, the direct distance between the base station antenna and the receiving antenna was approximately 74 m, indicating that the experimental environment corresponded to a non-line-of-sight (NLoS) condition due to building obstruction. The relative positions of the base station antenna and receiving antenna are illustrated in Fig. 3.

The measurements were carried out on five separate days during the academic term, from 9:00 to 19:00 each day. For pedestrian counting, we applied the YOLO11 object detection algorithm to detect pedestrian positions. Based on the detection results, the number of pedestrians was calculated for each of the three predefined counting areas. Given that the traversal time of the smallest counting area was approximately 2 seconds, a 2-second pedestrian count window was adopted for pedestrian counting.

## III. CORRELATION ANALYSIS BETWEEN PEDESTRIAN COUNT AND RSRP VARIANCE

We analyzed the relationship between the pedestrian count on the main street and the RSRP variance. Since the pedestrian counts were aggregated over a 2-second window, the RSRP variance was also averaged over 2 seconds to ensure temporal alignment. We call this value RSRP-Var($\Delta t$, 2s), where $\Delta t$ represents the variance window size used to calculate the RSRP variance, and the subsequent averaging is performed over 2 seconds. The correlation between RSRP-Var($\Delta t$, 2s) and the pedestrian count was calculated using Spearman's rank correlation coefficient $\rho$. In this context, a sample corresponds to a pair consisting of the RSRP-Var($\Delta t$, 2s) with a 1-second shift and the corresponding pedestrian count obtained from YOLO11. Thus, each sample represents an observation aligned at 1-second intervals, and $N$ denotes the total number of such pairs within a single measurement day. Spearman's rank correlation coefficient is defined as

$$\rho = 1 - \frac{6\sum_{i=1}^{N} d_i^2}{N(N^2-1)} \qquad (1)$$

where $d_i$ denotes the difference between the ranks of the $i$-th pair of samples.

Fig. 4 shows the averaged results across five measurement days for each feature RSRP-Var($\Delta t$, 2s) with different variance window sizes $\Delta t$. In the experiment, it was observed that larger counting areas tended to exhibit higher correlations. The variance window size $\Delta t$ that yielded the highest correlation was 0.1 seconds for the large area, and 0.2 seconds for both the medium and small areas. Overall, the correlation coefficients ranged from 0.46 to 0.59, which indicates a moderate level of correlation in statistical terms. These results suggest that RSRP-Var(Δ*t*, 2s) is a promising indicator for estimating pedestrian counts on the main street.

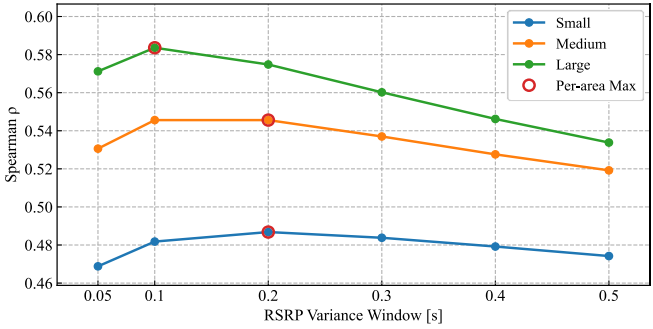

Fig.4 Spearman's ρ vs. Window size Δ*t* (5-days mean).

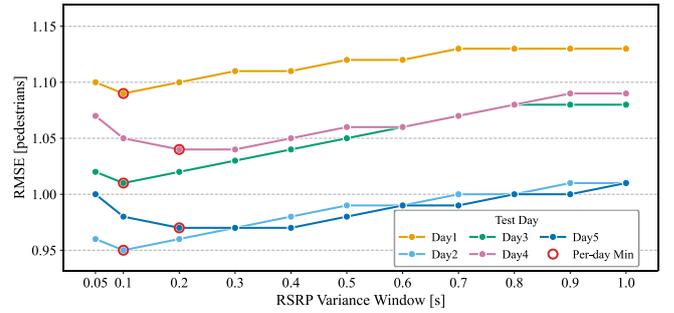

Fig.6 RMSE vs. Variance window size Δ*t*. (Medium area).

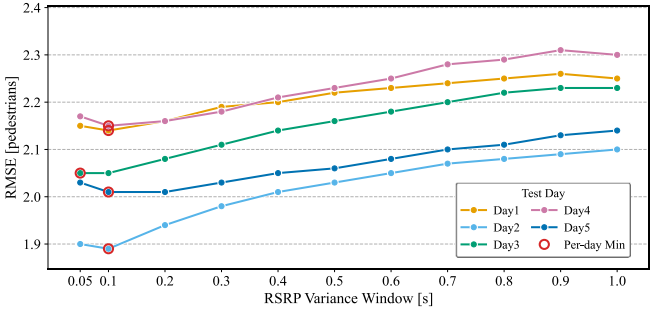

Fig.5 RMSE vs. Variance window size Δ*t* (Large area).

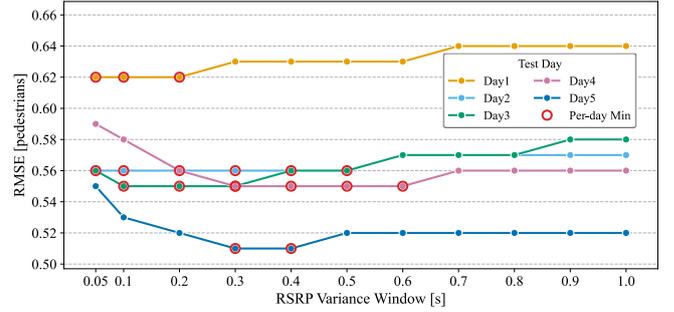

Fig.7 RMSE vs. Variance window size Δ*t* (Small area).

## IV. FEATURE ANALYSIS

### A. Definition of the Training Dataset

We use the CatBoost regression model for machine learning. The selected reason is that previous studies have demonstrated it outperforms other state-of-the-art methods such as, LightGBM, XGBoost, Random Forest, and Decision Tree, in terms of predictive accuracy [16]. In particular, CatBoost has been reported to achieve superior performance with respect to evaluation metrics such as the coefficient of determination ($R^2$) and the root mean squared error (RMSE), making it well-suited for the regression task of pedestrian count estimation in this study.

The training dataset was constructed as follows. The ground-truth pedestrian counts were obtained from YOLO11 using a two-second aggregation window, shifted in one-second increments. As the input features, we adopted RSRP-Var(Δ*t*, 2s), which was defined in the previous section. Specifically, the short-term variance of RSRP was first calculated with a variance window size of Δ*t*, and then averaged over a two-second interval to ensure temporal alignment with the pedestrian count.

Each sample was thus defined as a pair consisting of the RSRP-Var(Δ*t*, 2s) and the corresponding two-second pedestrian count. Measurements were conducted on the main street of the university campus over five separate days, from 9:00 a.m. to 7:00 p.m. For evaluation, four days of data were used for training, while the remaining one day was used for testing.

### B. Analysis of RSRP Variance Window Sizes

To investigate the optimal variance window size Δ*t* of RSRP-Var(Δ*t*, 2s) for crowd flow estimation, we evaluated a regression model trained with this single feature. Here, the single feature refers to the value of RSRP-Var(Δ*t*, 2s) computed for each sample, without incorporating any additional auxiliary features.

A sample was defined as a pair consisting of the RSRP-Var(Δ*t*, 2s) calculated with a one-second shift and the corresponding two-second pedestrian count obtained from YOLO11. In other words, each sample represents a paired observation of the RSRP variance feature and the pedestrian count for the same period. The analysis in this section was conducted using the data from a single measurement day.

The variance window size Δ*t* was varied incrementally from 0.05 to 1.0 seconds, and for each condition the CatBoost regression model was trained accordingly. The estimation error was quantified using the root mean squared error (RMSE), which is defined as

$$\text{RMSE} = \sqrt{\frac{1}{N}\sum_{i=1}^{N}(\hat{y}_i - y_i)^2} \qquad (2)$$

where $\hat{y}_i$ and $y_i$ denote the predicted and ground-truth pedestrian counts, respectively, and $N$ is the total number of samples within a single measurement day. Given that RMSE is sensitive to outliers, it provides a reliable indicator of errors associated with abrupt fluctuations in pedestrian counts. Therefore, RMSE was adopted as the primary evaluation metric in this study.

For training, data from four of the five measurement days were used, while the remaining day was for testing. The results for the large, medium, and small areas are presented in Figs. 5, 6, and 7, respectively. In the large area, four out of five test days recorded the minimum RMSE at a variance window size Δ*t* of 0.1 seconds, while the remaining day showed the minimum at 0.05 seconds, indicating that shorter variance window sizes were consistently effective. In the medium area, three test days yielded the minimum RMSE at 0.1 seconds and two days at 0.2 seconds, suggesting a relatively clear optimal range. In contrast, in the small area, the variance window size $\Delta t$ that minimized RMSE varied across test days, making it difficult to identify a unique optimal value. Within the same test day, differences in RMSE across variance window sizes were negligible, and no substantial variations in estimation performance were observed. However, when compared across different days, the variance window sizes $\Delta t$ yielding the minimum RMSE were distributed widely, ranging from 0.05 to 0.6 seconds, indicating a lack of stability. These findings confirm that, in the small area, the effect of the variance window size on estimation performance is limited, and it is challenging to determine a stable, unique optimal value.

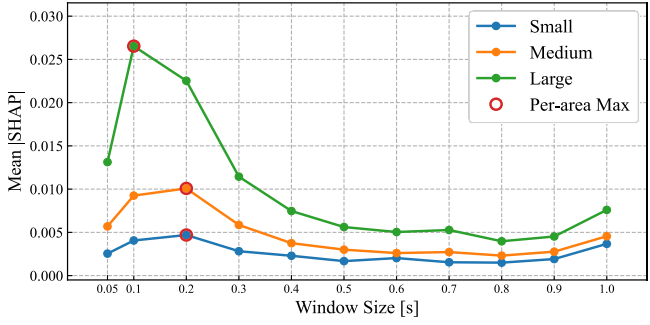

Fig.8  Mean |SHAP| vs. Variance window size $\Delta t$.

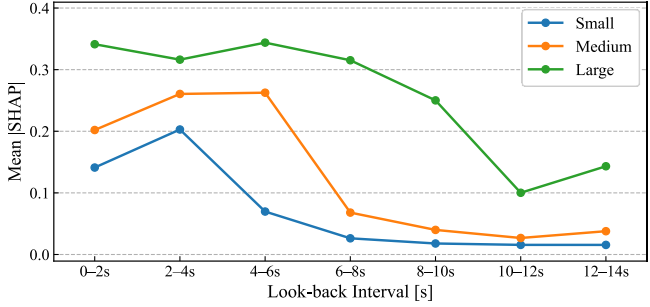

Fig.9  Mean |SHAP| vs. Look-back interval
(Variance window size $\Delta t$ = 0.2s).

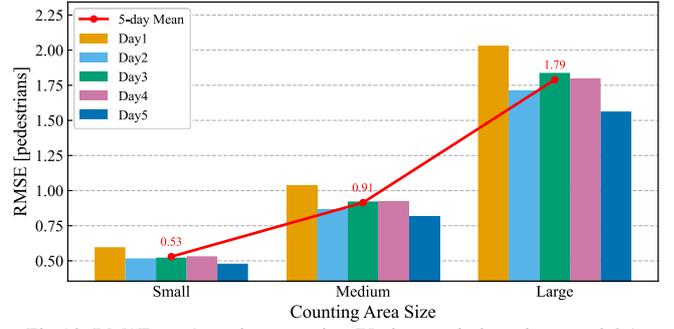

Fig.10  RMSE vs. Counting area size (Variance window size $\Delta t$ = 0.2s).

Next, to incorporate multiple variance window sizes simultaneously, we constructed feature vectors at each timestamp consisting of RSRP-Var($\Delta t$, 2s) for $\Delta t$ ranging from 0.05 to 1.0 seconds. A single CatBoost regression model was then trained on this joint feature set, enabling the simultaneous evaluation of the contributions of each variance window size $\Delta t$. Feature importance was quantified using SHapley Additive exPlanations (SHAP), Specifically, SHAP values were computed separately for each measurement day, and the mean absolute SHAP values were averaged across the five days to obtain representative contributions for each window size within each counting area. The resulting outcomes are presented in Fig. 8.

The analysis revealed that, across all areas, the contributions of $\Delta t$=0.1 s and $\Delta t$=0.2 s were particularly high, with $\Delta t$=0.1 s being the most effective in the large area, and $\Delta t$=0.2 s in the medium and small areas. These findings demonstrated that a variance window size of approximately 0.1–0.2 s is optimal for pedestrian count estimation.

Considering that the RSRP measurement interval was 0.01 s, this corresponds to variance values computed over 10–20 consecutive RSRP measurements. At $\Delta t$=0.05 s, the number of samples for calculating variance was insufficient, resulting in unstable variance values when no pedestrians were present. Conversely, at larger windows such as $\Delta t$=1.0 s, short-term multipath variations caused by pedestrian movement were excessively smoothed, thereby reducing estimation accuracy. Based on these observations, $\Delta t$=0.2 s was adopted as a representative and stable window size for subsequent evaluations.

*C. Contribution Analysis of Look-back Intervals*

To further clarify the temporal influence of RSRP-Var($\Delta t$, 2s) on crowd size estimation, a 14-second look-back period was divided into seven consecutive 2-second intervals: 0–2 s, 2–4 s, 4–6 s, 6–8 s, 8–10 s, 10–12 s, and 12–14 s. The RSRP-Var($\Delta t$, 2s) within each interval were used as features, so that multiple look-back intervals were simultaneously included in the input feature set. Since the ground-truth pedestrians were aggregated using a 2-second pedestrian count window, the same look-back period was applied to ensure temporal alignment between features and the ground-truth pedestrians. A CatBoost regression model was then trained with these features, and SHAP analysis was applied to quantify the contribution of each look-back interval. Data from four of the five measurement days were used for training, and the remaining day was reserved for testing. The mean SHAP values for each counting area are shown in Fig. 9.

As illustrated in Fig. 9, contributions remained high for a certain duration in all areas, but then exhibited markedly declines beyond specific intervals. In the small area, contributions were high in the 0–2 s and 2–4 s intervals, but decreased markedly after 4–6 s. In the medium area, contributions remained high up to 0–6 s, followed by a sharp reduction after 6–8 s. In the large area, contributions persisted until 0–8 s, then decreased to approximately half of their previous value at 8–10 s. These results indicate that contributions did not gradually decay but rather dropped steeply after particular intervals.

Furthermore, larger areas were found to sustain effective contributions over longer look-back intervals. This tendency can be attributed to the fact that pedestrians require more time to traverse larger areas; thus, the RSRP-Var($\Delta t$, 2s) from longer look-back periods carries more informative content for crowd size estimation.

Based on these findings, the effective look-back intervals were identified as 0–4 s for the small area, 0–6 s for the medium area, and 0–8 s for the large area. In the subsequent evaluations, these intervals were adopted as the primary features.

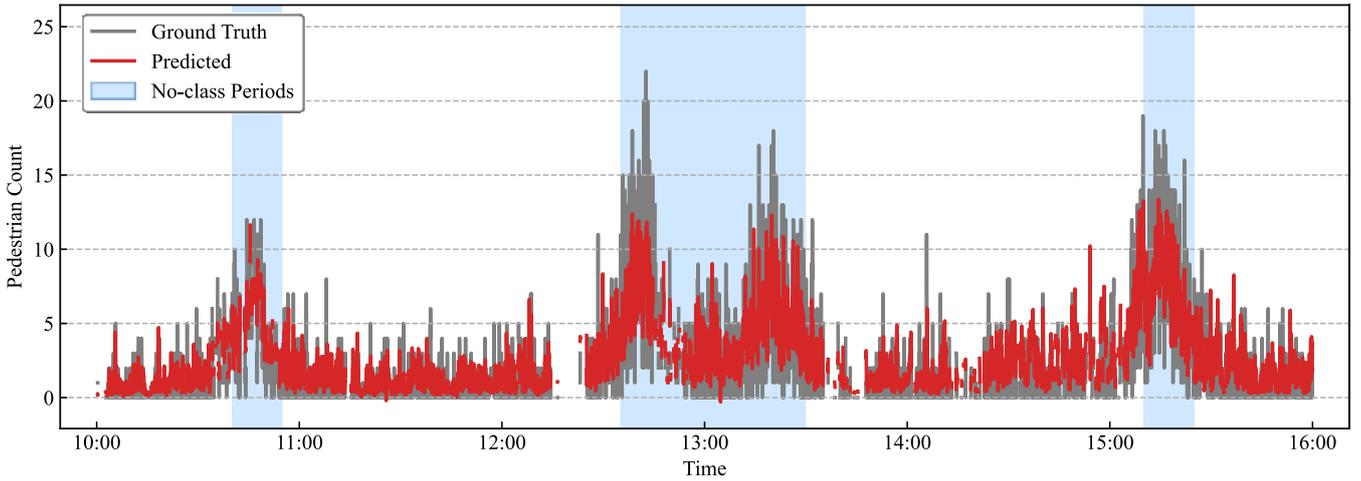

Fig.11 Ground truth vs. Predicted pedestrian count (Variance window size Δ$t$ = 0.2s, Large area, day5)

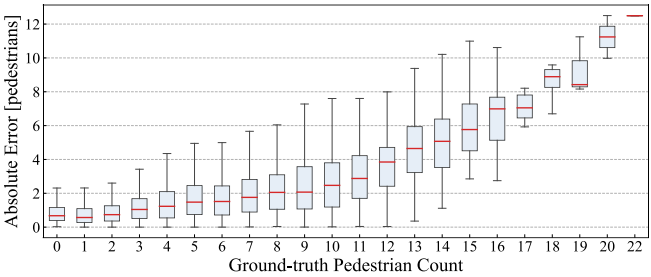

Fig.12 Absolute Error vs. Ground-truth Pedestrian Count
(Variance window size Δ$t$ = 0.2s, Large Area, day5).

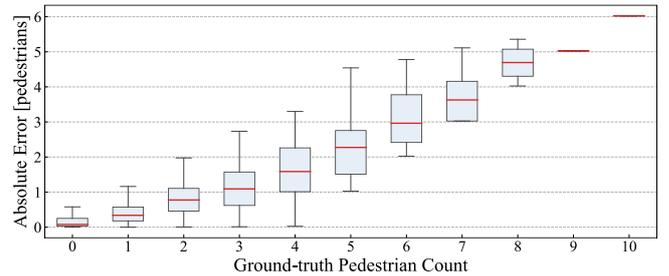

Fig.14 Absolute Error vs. Ground-truth Pedestrian Count
(Variance window size Δ$t$ = 0.2s, Small Area, day5).

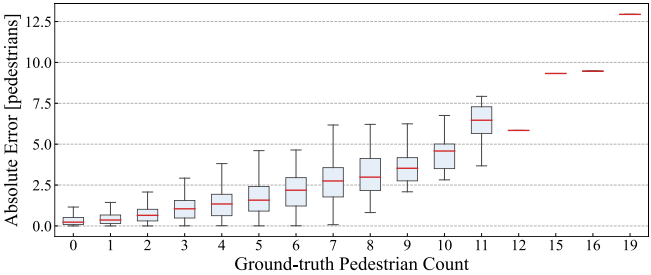

Fig.13 Absolute Error vs. Ground-truth Pedestrian Count
(Variance window size Δ$t$ = 0.2s, Medium Area, day5).

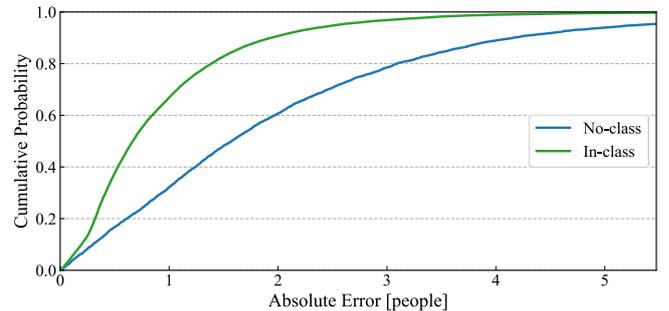

Fig.15 CDF of Absolute Error, In-class vs. No-class
(Variance window size Δ$t$ = 0.2s, Large Area, day5).

## V. Estimation Performance Evaluation

The performance of the crowd flow estimation was evaluated using the selected features. As evaluation metrics, both the RMSE and absolute error were employed.

Fig. 10 presents the RMSE results across five measurement days for each counting area. The results indicate that RMSE tended to increase as the counting area size became larger. This can be attributed to the fact that larger areas contain more pedestrians, leading to larger absolute errors. The mean RMSE values over five days were 0.53 persons for the small area, 0.91 persons for the medium area, and 1.79 persons for the large area, confirming that the RMSE remained within a range of two persons.

Fig. 11 illustrates the temporal variation of pedestrian counts on Day 5 in the large area as an example. Overall, the estimated values followed the temporal trends observed by the YOLO11. In particular, our analysis successfully reflected the temporal patterns of reduced flows during class hours and increased flows during breaks, demonstrating its capability to characterize variations over time. It should also be noted that the time-series graph includes missing intervals due to packet loss during data collection. However, during peak break times with more than ten pedestrians, the estimated values could not fully align with the ground-truth counts, resulting in errors of approximately ten pedestrians.

Fig. 12, 13, and 14 present the distribution of absolute errors by pedestrian count for the large, medium, and small areas on Day 5, using boxplots. In all areas, absolute errors were found to increase with the number of pedestrians. Moreover, as the pedestrian count increased, both the spread of the distributions and the number of outliers expanded rapidly, suggesting that under large-scale crowd conditions, pedestrian flows fluctuate substantially, making accurate pedestrian count estimation difficult.

Fig. 15 shows the cumulative distribution function (CDF) of estimation errors during class hours (In-class) and break times (No-class). Here, No-class refers to break periods without classes (10:40–10:55, 12:35–13:30, 15:10–15:25, 17:05–17:20), while In-class corresponds to the remaining class hours. The difference between In-class and No-class is the congestion on the street. During the No-class period, the number of pedestrians increases because they move from one classroom to another. During In-class periods, approximately 90% of the cases fell within an error of two persons, and about 60% within one person. The steep rise of the cumulative probability curve further confirmed stable estimation performance. In contrast, during No-class periods, approximately 90% of the cases involved errors involving four or fewer persons, and about 60% involved errors involving two or fewer persons. The gradual rise of the cumulative probability curve clearly indicated degraded estimation accuracy. These results demonstrate that during periods of low pedestrian density, estimation errors remain small and stable. In contrast, during break times, when large and abrupt fluctuations in pedestrian flow occur, errors increase and accuracy decreases.

Overall, our analysis was able to estimate the temporal variation of pedestrian flows on the main street with reasonable accuracy. Although estimation performance decreased during peak periods with large crowds, our method generally successfully tracked daily pedestrian flow patterns. These findings confirm the effectiveness of our analysis for capturing everyday crowd dynamics using commercial LTE. Future work includes expanding the dataset for large-scale crowds and designing a peak detection method to further improve estimation accuracy.

## VI. CONCLUSION

In this study, we conducted a field experiment on crowd flow estimation in an outdoor street environment using the variance of RSRP obtained from a commercial LTE base station. By associating camera-based pedestrian counts with RSRP variance and applying a CatBoost regression model together with SHAP analysis, we systematically evaluated the impact of feature design on estimation performance.

The analysis revealed that the optimal variance window size, $\Delta t$, of RSRP is 0.1–0.2 seconds for estimating pedestrian counts. Moreover, as the counting area was expanded, the effective look-back interval was extended, and the contributions of corresponding features increased. Regarding estimation performance, the estimation error as measured by RMSE, remained within approximately two persons' range even for the large area of 66.9 m². On the other hand, during peak periods with large crowds, estimation errors increased and accuracy degraded.

The contributions of this study can be summarized in two aspects. First, this work provides the first quantitative demonstration worldwide of outdoor crowd flow estimation using RSRP obtained from an operational commercial LTE base station. Second, it systematically clarifies how the variance window size and the counting area size influence estimation performance, thereby revealing the characteristic behavior of these features. These outcomes highlight crowd flow estimation as a promising use case for ISAC by extending existing LTE infrastructure to sensing and demonstrating its feasibility.

As future directions, we propose introducing peak-aware feature design to address degraded performance under large crowd conditions, integrating additional communication signal information, such as CSI or multi-cell RSRP, and conducting experimental validation in diverse urban environments. Through these extensions, this research is expected to further contribute to smart city services and next-generation urban management systems powered by ISAC.